\documentstyle[twoside,fleqn,espcrc2]{article}
\title{{Black Holes in Matrix Theory}}

\author{Miao Li {\sl and} Emil Martinec
\address{\it Enrico Fermi Institute and Dept. of Physics,\\
University of Chicago, 5640 S. Ellis Ave., Chicago, IL 60637 USA}}

\begin{document}


\def\pref#1{(\ref{#1})}

\def\ie{{i.e.}}
\def\eg{{e.g.}}
\def\cf{{c.f.}}
\def\etal{{et.al.}}
\def\etc{{etc.}}

\def\inbar{\,\vrule height1.5ex width.4pt depth0pt}
\def\IR{\relax{\rm I\kern-.18em R}}
\def\IC{\relax\hbox{$\inbar\kern-.3em{\rm C}$}}
\def\IH{\relax{\rm I\kern-.18em H}}
\def\IO{\relax\hbox{$\inbar\kern-.3em{\rm O}$}}
\def\IK{\relax{\rm I\kern-.18em K}}
\def\IP{\relax{\rm I\kern-.18em P}}
\def\Z{{\bf Z}}
\def\Pone{{\IC\rm P^1}}
\def\One{{1\hskip -3pt {\rm l}}}

\def\beq{\begin{equation}}
\def\eeq{\end{equation}}

\def\sst{\scriptstyle}
\def\tst#1{{\textstyle #1}}
\def\frac#1#2{{#1\over#2}}
\def\coeff#1#2{{\textstyle{#1\over #2}}}
\def\half{\frac12}
\def\hf{{\textstyle\half}}

\def\sdtimes{\mathbin{\hbox{\hskip2pt\vrule height 4.1pt depth -.3pt width
.25pt \hskip-2pt$\times$}}}
\def\bra#1{\left\langle #1\right|}
\def\ket#1{\left| #1\right\rangle}
\def\vev#1{\left\langle #1 \right\rangle}
\def\det{{\rm det}}
\def\tr{{\rm tr}}
\def\mod{{\rm mod}}
\def\sinh{{\rm sinh}}
\def\cosh{{\rm cosh}}
\def\sgn{{\rm sgn}}
\def\det{{\rm det}}
\def\exp{{\rm exp}}
\def\sh{{\rm sh}}
\def\ch{{\rm ch}}
\def\d{\partial}

\def\np{{\it Nucl. Phys. }}
\def\npb{{\it Nucl. Phys. }}
\def\pl{{\it Phys. Lett. }}
\def\plb{{\it Phys. Lett. }}
\def\pr{{\it Phys. Rev. }}
\def\prd{{\it Phys. Rev. }}
\def\ap{{\it Ann. Phys., NY }}
\def\prl{{\it Phys. Rev. Lett. }}
\def\mpl{{\it Mod. Phys. Lett. }}
\def\cmp{{\it Comm. Math. Phys. }}
\def\grg{{\it Gen. Rel. and Grav. }}
\def\cqg{{\it Class. Quant. Grav. }}
\def\ijmp{{\it Int. J. Mod. Phys. }}
\def\jmp{{\it J. Math. Phys. }}
\def\nextline{\hfil\break}
\catcode`\@=11
\def\slash#1{\mathord{\mathpalette\c@ncel{#1}}}
\overfullrule=0pt
\def\AA{{\cal A}}
\def\BB{{\cal B}}
\def\CC{{\cal C}}
\def\DD{{\cal D}}
\def\EE{{\cal E}}
\def\FF{{\cal F}}
\def\GG{{\cal G}}
\def\HH{{\cal H}}
\def\II{{\cal I}}
\def\JJ{{\cal J}}
\def\KK{{\cal K}}
\def\LL{{\cal L}}
\def\MM{{\cal M}}
\def\NN{{\cal N}}
\def\OO{{\cal O}}
\def\PP{{\cal P}}
\def\QQ{{\cal Q}}
\def\RR{{\cal R}}
\def\SS{{\cal S}}
\def\TT{{\cal T}}
\def\UU{{\cal U}}
\def\VV{{\cal V}}
\def\WW{{\cal W}}
\def\XX{{\cal X}}
\def\YY{{\cal Y}}
\def\ZZ{{\cal Z}}
\def\lam{\lambda}
\def\eps{\epsilon}
\def\vareps{\varepsilon}
\def\underrel#1\over#2{\mathrel{\mathop{\kern\z@#1}\limits_{#2}}}
\def\lapprox{{\underrel{\scriptstyle<}\over\sim}}
\def\lessapprox{{\buildrel{<}\over{\scriptstyle\sim}}}
\catcode`\@=12
%
\def\lpl{\ell_{p}}
\def\aleff{{\alpha'_{\rm eff}}}
\def\gym{g_{\scriptscriptstyle\rm YM}}
\def\bmatrix#1{\left[\matrix{#1}\right]}

\begin{abstract}
We review recent progress in understanding black hole structure
and dynamics via matrix theory.
Talk by the second author, presented at Strings '97 
(Amsterdam, June 16-20, 1997).
\end{abstract}

\maketitle

\section{Introduction}

Matrix theory \cite{bfss}
appears to capture a remarkable amount of
the nonperturbative structure of string/M-theory.  At generic
values of the moduli, most of the localized states of the latter 
are black holes.  One would like to see how matrix theory encodes
their basic properties -- geometry, dynamics of
test particles, thermodynamics, decay via the Hawking process,
etc.  First steps in this direction have recently been taken
\cite{lmone}-\cite{lmtwo}; 
our purpose here is to summarize the results
(with a few extensions), as well as to argue for a general picture
of black hole structure.  

We begin in section 2 with a discussion of the geometry we
are trying to reproduce -- classical eleven-dimensional
supergravity.  The focus is on five-dimensional (near)BPS
black holes \cite{bhrefs}
(for a review, see \cite{maldathesis}), 
as these are best understood from the
standpoint of string theory.  We then point out a few
properties of supergravitons as probes of the geometry, and
indicate how matrix theory reproduces various asymptotic 
features of the metric.

Adapting a D-brane calculation of Maldacena \cite{maldaone},
we reproduce in section 3 the near-extremal black hole entropy
as the entropy of the noncritical string theory that
describes matrix theory on $T^5$ \cite{dvv,seiberg}.
Rephrasing the calculation in natural M-theory variables
renders the interpretation of many formulae transparent.  The 
basic theme will be that black hole thermodynamics is 
generalized supersymmetric Yang-Mills statistical mechanics, 
in the microcanonical ensemble.\footnote{Here we are using an
inclusive notion of SYM to mean any of the theories that serve
to describe matrix theory on various tori.}

Section 4 presents some speculations on the dynamics of black holes
in matrix theory, summarizes our results and lists a few directions
for further research.


\section{Black holes in supergravity}

The class of black holes and black strings we will consider
arises in 11d supergravity on $\IR^{5,1}\times T^5$ 
\cite{sugrabh}.
Take the coordinates of $\IR^{5,1}$ to be ($x^0,...,x^4;x^{10}$);
sometimes we write $x^1,...,x^4$ in spherical coordinates
$r,\Omega_3$.  The coordinates of $T^5$ will be ($x^5,...,x^9$).
A black string stretched along $x_{10}$
may be constructed as a bound state of fivebranes,
membranes, and gravitons of the 11d theory in the configuration
\beq
\bmatrix{.&6&7&8&9&10    \cr
           .&.&.&.&.&p_{10}\cr
           5&.&.&.&.&10    \cr }\ .
\label{config}
\eeq
Compactification of $x_{10}$ on a circle produces a black hole
in five-dimensional spacetime.
The generic (nonextremal) metric for such a configuration is 
\begin{eqnarray}
\lefteqn{ds^2= H_2^{1/3}H_5^{2/3}\times}\nonumber\\
	& &\bigl[H_2^{-1}  H_5^{-1}(-H_0^{-1}h\;dx_0^2
                + H_0 \widehat{dx_{10}}^2)\nonumber\\
        & &\ \ +H_2^{-1}dx_5^2 + H_5^{-1}(dx_6^2+...+dx_9^2)\nonumber\\
        & &\ \ \ \  +h^{-1} dr^2 + r^2 d\Omega_3^2\bigr]\ ,
\label{nonextreme}
\end{eqnarray}
where $H_i(r)$, $i=0,2,5$, and $h(r)$ are harmonic functions
\begin{eqnarray}
H_i&=&1+\frac{r_i^2}{r^2}\quad r_i^2=r_g^2\;\sinh^2\alpha_i \\
h&=&1-\frac{r_g^2}{r^2}\quad r_g={\rm grav.~(horizon)~radius}\ ,\nonumber
\end{eqnarray}
and $\widehat{dx_{10}}$ is a combination of $x_{10}$ and $x_0$
whose specific form will not be needed.  The extremal limit of
this metric involves $r_g\rightarrow0$, $\alpha_i\rightarrow\infty$,
with the charge radii $r_i$ held fixed:
\begin{eqnarray}
\lefteqn{ds^2 =H_2^{1/3}H_5^{2/3}\times}\nonumber\\
	& & \bigl[H_2^{-1}H_5^{-1}(du\;dv+(H_0-1)du^2)\nonumber\\
	& &\ +H_2^{-1} dx_5^2+H_5^{-1}(dx_6^2+...+dx_9^2)\nonumber\\
	& &\ \ +dr^2+r^2d\Omega_3^2\bigr]\ ,
\label{extreme}
\end{eqnarray}
where $u,v=x_0\pm x_{10}$.  We will soon take $v$ to be light front
time in the infinite momentum frame (IMF) of the matrix theory construction.
The motion of a supergraviton probe (or D0-brane in the language
of matrix theory) in the extreme black hole background
is governed by the Laplacian
\begin{eqnarray}
\lefteqn{\Delta=H_2H_5[\d_u\d_v-(H_0-1)\d_v^2]}\nonumber\\
	& &\ +(\d_1^2+...+\d_4^2)\nonumber\\
	& &\ \ +H_2\d_5^2+H_5(\d_6^2+...+\d_9^2)\nonumber\\
	& &\ \ \ +({\rm connection~terms})\ .
\label{laplacian}
\end{eqnarray}
For a probe of the causal structure of the black string, 
we want a massless particle, therefore we consider waves
independent of the internal coordinates on the $T^5$.
One can further choose special polarizations of supergraviton
(\eg\ the 567 component of $A_{MNP}$) such that the connection
term vanishes.  Wavepackets of this sort behave as if they are
scalars, hence travel light cones of the 11d metric -- 
precisely the probes we want.  A wavepacket $\psi$ with 
$\d_u\psi\gg\d_v\psi$ ($p_+\gg p_-=E_{LC}$)
comoves with the wave bound to the black string and is therefore
near-BPS.  In a sense, it adiabatically approaches and crosses the
horizon.  The wave equation $\Delta\psi=0$ reduces to
\beq
n^2(r)\omega_{\rm eff}^2\psi={\vec \d}^2\psi\ ,
\label{probe}
\eeq
where ${\vec\d}^2=\d_1^2+...+\d_4^2$, 
$\omega_{\rm eff}^2\psi\approx\d_u\d_v\psi$, and $n(r)=H_2H_5$.
Thus the effective dynamics of wavepackets is equivalent to 
geometric optics with the spatially dependent index of refraction $n(r)$.
Since the index increases as the radius decreases, null 
geodesics are focussed onto the string (gravitational lensing).

In the next section we will exhibit a matrix theory configuration
for near-extremal black strings/holes of the type \pref{nonextreme}.
The effective dynamics maps onto an equivalent D-brane system
(5+1d SYM), although the interpretation is rather different.
One of the most important distinctions is that, whereas the D-brane 
moduli space describes slowly moving massive particles, the matrix model
moduli space describes the transverse motion of {\it massless}
particles (the supergravitons) -- thus giving direct information
about the causal structure.

Calculations in this D-brane system \cite{dkps,bfss,dps,maldatwo}
reproduce all $1/r^2$ terms in the near-extremal geometry.
Curved space geometry is a {\it low-energy} approximation in matrix
theory.  The dynamics of the probe-black hole system in $\IR^{5,1}$
is described by matrix Higgs fields in the 
D-brane SYM on $T^5$ ($x_5,...,x_9$)
\beq
\vec\Phi=\bmatrix{\vec X_{BH}&\vec Y\cr
		  \vec Y^{\dagger}&\vec X_{\rm probe}}\ ,
\eeq
where the vector components run over ($x_1,...,x_4$).
The $\vec Y$ variables are massive, schematically
$V_{\rm eff}=[\Phi^i,\Phi^j]^2\sim 
|\vec X_{BH}-\vec X_{\rm probe}|^2|\vec Y|^2\equiv r^2|\vec Y|^2$.
Integrating out $\vec Y$ yields an effective action that
is an expansion in powers of derivatives of the light fields $X$ 
and inverse powers of $r$.
All the $(r_i/r)^2$ terms in the asymptotic metric come
from integrating out $\vec Y$ 
at one loop \cite{dkps,bfss,dps,maldatwo}.\footnote{The interaction 
with the gravitational wave $H_0$ bound to the string is 
none other than the
$(\vec v_{\rm probe}-\vec v_{\rm BH~wave})^4/r^7$ 
`Coulomb' interaction of matrix theory \cite{bfss};
the $r^{-7}$ becomes $r^{-2}$ when smeared over the internal $T^5$.}
Thus, in matrix theory $n^2(r)\sim 1+\frac{r_2^2}{r^2}+\frac{r_5^2}{r^2}$
{\it is} an optical index -- it is generated by the spatially dependent
vacuum polarization whose effect is a spatially varying
dielectric function seen by the light degrees of freedom.
However, this correspondence makes it clear that the light-cone
structure of spacetime is indeed only a low-energy approximation
in matrix theory.  This structure is defined by the trajectories
of massless particles, such as the probe wavepackets considered above.
A supergraviton probe will only follow the classical black hole
light cones in this (moduli space) approximation, \ie\ when the
$Y$ variables are sufficiently heavy that they may be
consistently integrated out.  Near the black hole (the `stretched horizon'
in black hole physics \cite{stretchor}; the `stadium region' in 
D-brane terminology \cite{dkps}), this approximation breaks down.
Thus in the Schwarzschild-type coordinate frame intrinsic to
the matrix description, the classical causal structure
and classical notions of information propagation only make
sense sufficiently far from the black hole (or string in the
present case).  Near the hole, spacetime literally becomes
nonabelian, and the light cones are not meaningful (at least
in this coordinate frame).

Of course, the $1/r^2$ terms in the metric are all {\it required}
by Gauss' law -- they represent the energy of gauge fluxes
carrying the BPS or Noether charges ($Q_0,Q_2,Q_5,E$) at infinity.
It is more a relief than a triumph for matrix theory to reproduce them.
Subleading terms are allowed to depend sensitively 
on nonuniversal details -- the wavefunction of the black hole and
probe degrees of freedom, form factors for their scattering, etc.
It may be that the precise black hole geometry is reproduced only
in the large N limit, with all these effects taken into account.
In other words, we need to understand how the Einstein equations
come out of matrix theory.  It is encouraging that the behavior
of supergraviton probes \pref{probe} has the same form that one
would expect to get from the abelianized moduli space approximation
to matrix theory -- the classical geometry acts as an `optical medium'
whose optical index bends the trajectories of probes.

\section{Matrix black hole thermodynamics}

\noindent{\bf The entropy:}
To discuss thermodynamic properties, we will postulate 1) that
there exists some theory (perhaps of noncritical strings 
\cite{dvv,seiberg}) whose large N dynamics formulates matrix theory
on $T^5$; and 2) this theory has a regime in which it is well-approximated
by 5+1d Yang-Mills with 16 supersymmetries.  One can think of 
this auxilliary theory heuristically as the T-dual of
the original matrix theory on all five circles of $T^5$:
\begin{eqnarray}
{\rm \sst graviton}\sim D0&\longrightarrow& D5 \nonumber\\
{\rm \sst long.~fivebrane}\sim D4&\longrightarrow & D1 \\
{\rm \sst long.~membrane}\sim{\rm IIA~\sst string}&\longrightarrow&
	{\rm \sst momentum} \ .\nonumber
\end{eqnarray}
Thus the three charges correspond, respectively, to the rank $N$
of the gauge group, the instanton number,
and the field momentum in the SYM theory.
This maps the system to a well-studied type IIB D-brane system
(for a review, see \cite{maldathesis}).  Finite N in matrix theory
is supposed to be related to compactification of the 
longitudinal \cite{bfss} or light-front \cite{dlcq} coordinate 
on a circle of radius $R$.  In matrix theory, branes' wrapping/momenta
are SYM fluxes:
\begin{eqnarray}
	q_i&=&\int \tr F_{0i}=
		\left({\sst{\rm graviton~KK~charge}\atop
			{\sst\rm along~}{\sst x_i}}\right)
  \nonumber\\
	m_{ij}&=&\int\tr F_{ij}=
		\left({\sst{\rm membrane~wrapping~}\atop
			{\sst x_i x_j~{\rm cycle}}}\right)
  \nonumber\\
	m_{+i}&=&\int T_{0i}=
		\left({\sst{\rm longitudinal~membrane}\atop
			\sst {\rm wrapping~}x_i x_{10}}\right)
  \nonumber\\
	f_i&=&\int(F\wedge F)_i=\int\eps_{ijklm}F_{jk}F_{lm}\nonumber\\
		&=&\left({\sst{\rm longitudinal~fivebrane~along~}\atop
			\sst x_j x_k x_l x_m x_{10}}\right)
  \ .
\label{fluxes}
\end{eqnarray}
Note that the instanton is a solitonic string in 5+1d, and $f_i$
is its wrapping along the $x_i$ cycle.  The configuration
\pref{config} corresponds to exciting $m_{+5}$ and $f_5$
in the SYM theory.

Horowitz, Maldacena, and Strominger \cite{hms} proposed
an identification of the parameters of the classical geometry with
`brane charges'
\begin{eqnarray}
N_{0,\bar 0}\frac{\lpl}{R}&=&\frac{VR_5R}{4\lpl^8}r_g^2e^{\pm2\alpha_0}
	\equiv E_{0,\bar0}\nonumber\\
N_{2,\bar 2}\frac{R_5R}{\lpl^2}&=&\frac{VR_5R}{4\lpl^8}r_g^2e^{\pm2\alpha_2}
	\equiv E_{2,\bar2}\nonumber\\
N_{5,\bar 5}\frac{VR}{\lpl^5}&=&\frac{VR_5R}{4\lpl^8}r_g^2e^{\pm2\alpha_5}
	\equiv E_{5,\bar5}\ ,
\label{branes}
\end{eqnarray}
where $V=R_6R_7R_8R_9$ is the volume of the $T^4$ spanned
by the fivebrane, and $E_i$, $E_{\bar i}$ are the contributions
of these `constituents' to the ADM energy
\beq
E_{ADM}=\sum_{i=0,2,5}(E_i+E_{\bar i})\ .
\eeq
We will argue below that the identifications \pref{branes}
are somewhat misleading, nevertheless they will serve our purpose
for the moment.  

In the classical supergravity solution,
one requires a balance of the pressures and tensions exerted by
the various branes on the internal $T^5$ in order to have a
nonsingular horizon at extremality; in terms of the quantities 
\pref{branes},
\beq
E_0E_{\bar0}=E_2E_{\bar2}=E_5E_{\bar5}\ .
\label{balance}
\eeq
We propose to interpret this as an equipartition of the
`invariant masses' among the different BPS charges, since
the first of these is the invariant mass of a 1d gas, and the
others are U-dual to such a quantity.  Note that $R$, $R_5$ and $V$
act in various combinations as chemical potentials for the
different branes; for instance, decreasing $R_5$ and increasing $V$
(with $R_5V$ held fixed in order to keep $\gym^2$ constant)
increases the proportion of 2-branes relative to 5-branes.
Finally, we take $N\sim N_0\gg N_{\bar0}$ in order to
match IMF dynamics.

Now consider a gas of instanton strings (anti)winding along $x_5$.
Instanton charge on a torus fractionalizes into N pieces,
leading to a much longer effective string with a tension reduced
by a factor 1/N.  When we excite these strings, they will 
quickly enter the Hagedorn phase because of 
their low tension\footnote{This phase is well-defined at fixed energy
(microcanonical ensemble).}.  All quantum numbers of the system --
energy, winding, momentum -- are carried by a single long string.
This is because the instanton string has by far the lightest
excitations in the gauge theory (due to its length being much longer
than the size of the $T^5$).  Thus, the black hole entropy 
is that of a single long string carrying all the 2-brane
and 5-brane energy and charges:
\begin{eqnarray}
\lpl H_{SYM}&=&\lpl E_{LC}=\lpl(E_{ADM}-\frac NR)\nonumber\\
	&=&(N_2+N_{\bar2})\frac{RR_5}{\lpl^2}\nonumber\\
	& &\quad +(N_5+N_{\bar5})\frac{RV}{\lpl^5}\nonumber\\
Q_2\equiv \lpl\PP&=&(N_2-N_{\bar2})\frac{RR_5}{\lpl^2}\nonumber\\
Q_5\equiv \lpl\WW&=&(N_5-N_{\bar5})\frac{RV}{\lpl^5}\ .
\label{charges}
\end{eqnarray}
The Virasoro constraints on the instanton string determine
its excitation level:
\beq
n_{L,R}=\aleff[E_{LC}^2-(\PP\pm\WW)^2]\ .
\label{levelno}
\eeq
Plugging in and using $T_{\rm eff}=(2\pi\aleff)^{-1}
=(4\pi^2/\gym^2N)=\frac{VR_5R^2}{2\pi N\lpl^9}$,
one finds the well-known answer \cite{hms,maldaone}
\begin{eqnarray}
S_{BH}&=&2\pi(\sqrt{n_L}+\sqrt{n_R})\nonumber\\
  &=&2\pi\sqrt{N}(\sqrt{N_2}+\sqrt{N_{\bar2}})\nonumber\\
	& &\quad\times(\sqrt{N_5}+\sqrt{N_{\bar5}})\ .
\label{sbh}
\end{eqnarray}
In the $R,N\rightarrow\infty$ limit,\footnote{With $N/R^2$ fixed
in order to have a finite entropy per unit length in the limit.} 
a graviton with any nonzero
transverse velocity can be boosted into the forward
direction (since the transverse momentum acts effectively as
a mass with respect to longitudinal boosts).  If matrix theory is
Lorentz covariant, in this limit there may be a sense in which {\it all}
nonextremalities are effectively turned on.\footnote{Note, however,
that when $R>r_g$ an instability develops \cite{gregory}, 
whereby the black string will become a black hole threaded by an extremal
string.  This suggests we keep $R$, $N$ finite of order the
size of the black hole in order to capture the relevant physics.}
Finally, the fact that we get the right answer independent of the moduli
of $T^5$ suggests that the above picture captures the thermodynamic
properties of the noncritical string of matrix theory on $T^5$
\cite{dvv,seiberg}.

There are a few simple generalizations of the above calculation.
First, one can turn on other BPS charges \cite{dvv}.  The general
extremal entropy is
\begin{eqnarray}
S_{BH}&=&2\pi\bigl[(Nf_i+\hf(m\wedge m)_i)\nonumber\\
	& &\quad\times(m_{i+}-q^jm_{ij}/N)\bigr]^{1/2}\ .
\label{gensbh}
\end{eqnarray}
Halyo \cite{halyo} has considered the (near)extremal black hole
obtained from transverse rather than longitudinal gravitons
and membranes (the second term rather than the first in the last
factor).  This should correspond to turning off the momentum charge
carried by the Hagedorn string, replacing 
its contribution to the energy by that
of the transverse zero- and two-brane charges that are
turned on instead.  Second, the considerations of \cite{maldaone}
included Hagedorn strings with angular momentum; it is straightforward
to translate the results to the context of matrix theory
to obtain the entropy of spinning matrix black holes.
Third, nothing in the entropy calculation required 
$\vec\PP||\vec\WW$; this simply corresponds to choosing 
orthogonally intersecting membranes and fivebranes.
Allowing $\vec\PP$ to make an angle $\zeta$ with respect to $\vec\WW$
means that there is a component of the membrane charge along
the fivebrane -- the branes intersect at angles.  Repeating
the above exercise, one finds \eg\
\begin{eqnarray}
n_{L,R}&=&\aleff[E_{LC}^2-(\PP_5\pm\WW_5)^2-\PP_9^2]\nonumber\\
	&=&(\frac{\aleff}{\lpl^2})
		\biggl[\frac{VRR_5}{4\lpl^8}\;r_g^2\biggr]^2
\label{nonextent}\\
	& &\hskip -2.2cm\times\bigl[\cos^2\zeta\;\ch^2(\alpha_2\pm\alpha_5)
	+\sin^2\zeta\;\ch^2(\alpha_2+\alpha_5)\bigr]\ .\nonumber
\end{eqnarray}
In the extremal limit $\alpha_2,\alpha_5\rightarrow\infty$
the entropy agrees with known results \cite{cocv};
the nonextremal entropy has not been computed in supergravity,
and it would be interesting to see if it matches \pref{nonextent}.

\noindent{\bf Interpreting the perturbation from extremality:}
It was claimed above that the identification \pref{branes}
is somewhat misleading.  To see this, let us consider
the excitations of the instanton string.  Its transverse
oscillators (in Green-Schwarz formalism) consist of
$X^i$, $S_L^{\alpha a}$, and $S_R^{\dot\beta \dot b}$,
where $\alpha,\dot\beta$ are spinor indices of the SO(4)
transverse to the $T^5$ (the R-symmetry of the 5+1d SYM),
and $i,a,\dot b$ are vectors and spinors of the SO(4)
little group in 5+1d.  The center of mass mode of the string
has the quantum numbers of the vector multiplet (the string
we are considering is the T-dual of the one in \cite{dvv}),
and one can interpret the oscillator modes as coupling
to fluctuations of the SYM theory.  Therefore local oscillations
along the Hagedorn string are fluctuations of the
Hagedorn gas, equivalent to {\it all types} of local
fluctuations of SYM.  This picture may carry over 
largely unmodified to the full noncritical interacting matrix
string theory of \cite{dvv,seiberg}; there, the SYM `particles'
are microscopic noncritical strings, and the Hagedorn string
is simply a macroscopic string cut from the same cloth.
Dumping a lot of energy into the system forces it into
the Hagedorn phase, and the soft excitations of the Hagedorn
string indeed couple to {\it all} quantum numbers of the theory.
This explains why \pref{sbh},\pref{gensbh} are valid 
over all of the parameter space of the 5+1d matrix theory.

We now see why the identification \pref{branes} is not
the full story.  The Hagedorn gas contains fluctuations of
all possible quantum numbers \pref{fluxes},
not just those which have an expectation value such as
$m_{+5}$ and $f_5$.  Rather, the energy above extremality is
equipartitioned into all types of branes/antibranes
according to their energy cost per quantum.
What were called $N_i$, $N_{\bar i}$, $i=0,2,5$ in \pref{branes},
are simply a characterization of the state of excitation of the 
Hagedorn string.  They
were actually determined by the ADM charges $E_{ADM}$, $Q_0$, $Q_2$,
$Q_5$, and the two `pressure balance' conditions 
\pref{balance}.\footnote{The membrane/fivebrane pressure balance
can be motivated by extremizing the entropy of the
Hagedorn string \cite{lmtwo}; the remaining pressure balance
will require an interpretation within matrix theory of the
factor of $N_{\bar0}$ appearing in \pref{balance}.
Perhaps one can get both pressure balance conditions
by turning on other charges as in the discussion after \pref{gensbh}.}

\section{Discussion and speculations}

\noindent{\bf The infalling probe:}
We have seen that matrix theory encodes the key features
of 5+1d black holes -- their leading asymptotic geometry
and their density of states.  Whether the rest of the structure
is present is tantamount to an understanding of how general relativity
appears as the effective theory.  It is at least encouraging
that, in supergravity,
the motion of probes in the black hole background
can be recast in the form of geometric optics.  
In matrix theory, this is just
what one gets from the vacuum polarization effects of the
heavy matrix degrees of freedom.  The issue is whether
these effects reproduce the right `optical index'.

A legitimate question to ask of matrix theory concerns the
fate of an infalling probe.  In the exterior static coordinates
intrinsic to the infinite momentum frame of matrix theory,\footnote{This
is because the optical index blows up near the horizon,
causing the probe wave to infinitely slow down, as one expects
in static coordinates.  The divergence is fake, however, merely
reflecting the failure of the approximation made in integrating
out off-diagonal matrix elements which are becoming light there.}
when the probe reaches the `stretched horizon' of the black hole,
it dissolves into the Hagedorn gas of the black hole.
In terms of matrices, the evolution has the schematic form
\beq
\bmatrix{{\bf BH}&0\cr 0&{\rm probe}}
	\longrightarrow\biggl[~{\bf BH^*~}\biggr]\ ,
\eeq
where $\bf BH^*$ denotes an excited black hole.  To describe the
proper motion of the probe one would like to at least
approximately diagonalize the probe's collective field theory.
This would give an approximate evolution looking like
\beq
\bmatrix{{\bf BH}&0\cr 0&{\rm probe}}
        \longrightarrow \bmatrix{{\bf BH'}&0\cr 0&{\rm probe'}}\ ,
\eeq
at least until one gets close to the singularity.
Such a rediagonalization should involve passing to infalling
coordinates, for instance by performing a sequence of boosts 
to keep the probe in its instantaneous rest frame.  
In general relativity, the
coordinate transformation that results has the form
\begin{eqnarray}
u&\rightarrow &U\sim\coeff{1}{a} e^{au}\nonumber\\
v&\rightarrow &V\sim v-a(\coeff{b}{r})^2\nonumber\\
r&\rightarrow &W\sim e^{-au}\cdot(\coeff {r}{b})\ .
\end{eqnarray}
The first of these undoes the exponential redshift of static
coordinates near the horizon, the second is a large shift of 
light-front time, and the last rewrites the radial variable
as a kind of `tortoise' coordinate.  The black hole interior
has $U<0$, $W<0$.  But this is an analytic continuation of
the IMF description; $u$ is the conjugate variable to $p_-\sim N$,
so the meaning of continuing past $u=\infty$ is unclear.
Similarly, $W<0$ may involve a continuation to complex eigenvalues
of ${\vec X}_{\rm probe}$.  Even so, it seems clear that the
sequence of boosts required to keep the probe in its rest frame
involves boosts which mix light-front time and ${\vec X}_{\rm probe}$;
therefore, the probe proper time is a noncommutative (matrix)
variable, and the evolution equation of the probe with
respect to its proper time involves a moduli space approximation
in this time coordinate.  One might imagine that the classical 
singularity of the black hole is simply a reflection of the
breakdown of this moduli space approximation in the time
direction.  It is also interesting to note that the boosts 
involved are matrix transformations; it is possible that 
observables appropriate to asymptotic observers and observables
measured in the infalling frame will not commute,
leading to a form of black hole complementarity \cite{stretchor}.

\noindent{\bf Directions for further research:}
It is important to expand the lexicon of translations 
between matrix theoretic and gravitational quantities, especially the
gross geometrical features of the region near the black hole.
It should be possible to calculate the properties of Hawking radiation
in the matrix theory approach.  Near extremality, the
Hawking temperature and the temperature of the Hagedorn gas
coincide \cite{maldaone}.  One can relate the Hawking temperature
to the temperature of the Hagedorn string arbitrarily far from
extremality \cite{lmtwo}
\beq
\beta_H=\beta_{str}[4(\aleff)^{1/2} E_{LC}]\ .
\eeq
It would be nice to understand the factor of proportionality.
There are also similar relations which need a proper explanation,
for instance the formula
\beq
(n_Ln_R)^{1/2}=\frac NR\;E_{LC}\;r_g^2
\eeq
which relates the Hagedorn string's excitation to the
gravitational radius of the black hole.\footnote{In principle,
$r_g^2\sim\vev{{\vec X}_\perp^2}$; the excitations of ${\vec X}_\perp$
are R-R fluctuations of the Hagedorn string and are determined
by equipartition.  This is related to the fact that absorption of
higher angular momenta proceeds via the worldsheet fermions
\cite{angular}.}
Larsen \cite{larsen} has related $n_L-n_R$ to
the area of the inner horizon.
How are these reflected in the dynamics of probes?

Perhaps the main lesson to be drawn from the above analysis
is that black hole thermodynamics becomes conventional statistical
mechanics in matrix theory.  Thus supergravity gives a whole
host of predictions -- for the (2,0) field theory governing
matrix theory on $T^4$, for $\NN$=4 SYM on $T^3$, and so on.
The thermodynamic properties of these higher-dimensional
black holes (\cf\ \cite{sugrabh}) are currently not
understood in string theory; there is no weak coupling limit
where the horizon is nonsingular near extremality.  Nevertheless,
one might hope to directly study the nonextremal black hole
via matrix theory.  The challenge is to find the relevant
degrees of freedom and to understand their behavior.

If we are successful, we should be able to explain the black
hole correspondence principle \cite{hp}.  We saw that the moduli
of the internal torus act as chemical potentials, altering
the balance of degrees of freedom in the equilibrium SYM
ensemble.  At small $R_5$, membrane/antimembrane excitations
dominate; at small $V$, fivebrane/antifivebrane excitations
are more prominent.  Shrinking any circle of
the torus to sub-Planckian size, one will recover
the matrix description of 
black holes as bound states of perturbative strings and D-branes.
It is implicit in \cite{matstr}
that the level density is $o(\exp[M])$ in the perturbative
string regime.  The black hole level density is 
$o(\exp[M^{\frac{d-2}{d-3}}]$, so there will have
to be a substantial crossover in the SYM level density
as a function of $R$.
The correspondence principle is a statement that there are
no phase transitions in this crossover region,
so that the spectra of black holes and of string states
match smoothly onto one another.

Perhaps the simplest example of this phenomenon occurs in
matrix 1+1d SYM on $S^1$.  The flux $\oint T_{01}$ (a momentum
mode of the SYM) represents a longitudinally wrapped membrane.
At large radius and far from extremality, the SYM statistical
mechanics describes a black string; at small radius,
an excited fundamental IIA string emerges, stretched
across the longitudinal direction.  This situation is
currently under investigation \cite{lmwip}.

\vskip .5cm
\noindent {\bf Acknowledgements:} 
E.M. thanks the Aspen Center for Physics for hospitality during
the preparation of the manuscript.
This work is supported in part by funds provided by the DOE under
grant No. DE-FG02-90ER-40560.


\begin{thebibliography}{99}
\small

\bibitem{bfss}{T. Banks, W. Fischler, S.H. Shenker and L. Susskind,
hep-th/9610043.}%

\bibitem{lmone}{M. Li and E. Martinec, hep-th/9703211.}%

\bibitem{lmtwo}{M. Li and E. Martinec, hep-th/9704134.}%

\bibitem{dvv}{R. Dijkgraaf, E. Verlinde, and H. Verlinde, hep-th/9704018.}%

\bibitem{bhrefs}
A. Strominger and C. Vafa, hep-th/9601029, 
\plb {\bf 379} (1996) {99};
C. Callan and J. Maldacena, hep-th/9602043, 
\npb {\bf 472} (1996) {591};
G. Horowitz and A. Strominger, hep-th/9602051, 
\prl {\bf 77} (1996) {2368};
S.R. Das and S.D. Mathur, hep-th/9606185, 
\npb {\bf 478} (1996) {561};
A. Dhar, G. Mandal, and S. Wadia, hep-th/9605234, 
\plb {\bf 388} (1996) {51};
J. Maldacena and A. Strominger, hep-th/9609026, 
\prd {\bf 55} (1997) {861}.

\bibitem{maldathesis}
J. Maldacena, PhD thesis, hep-th/9607235.%

\bibitem{maldaone}
J. Maldacena, hep-th/9605016, \npb {\bf 477} (1996) {168}.

\bibitem{seiberg}
N. Seiberg, hep-th/9705221.

\bibitem{angular}
S. Mathur, hep-th/9704156; S. Gubser, hep-th/9704195.

\bibitem{larsen}
F. Larsen, hep-th/9702153,
\prd {\bf 56} (1997) {1005}.

\bibitem{sugrabh}
A.A. Tseytlin, hep-th/9604035, \npb {\bf 475} (1996) {149};
M. Cvetic and A. Tseytlin, hep-th/9606033, \npb {\bf 478} (1996) {181};
M.J. Duff, H. Lu, and C.N. Pope, hep-th/9604052, 
\plb {\bf 382} (1996) {73}.

\bibitem{dkps}
M. Douglas, D. Kabat, P. Pouliot, and S.H. Shenker,
hep-th/9608024, \npb {\bf 485} (1997) {85}.

\bibitem{dps}
M. Douglas, J. Polchinski and A. Strominger, hep-th/9703031.

\bibitem{maldatwo}
J. Maldacena, hep-th/9705053.

\bibitem{stretchor}
L. Susskind, L. Thorlacius, and J. Uglum,
hep-th/9306069, \prd {\bf 48} (1993) {3743};
G. 't Hooft, unpublished;
K. Schoutens, E. Verlinde, and H. Verlinde, hep-th/9401081;
Y.Kim, E. Verlinde, and H. Verlinde, hep-th/9502074, 
\prd {\bf 52} (1995) {7053}.

\bibitem{dlcq}
L. Susskind, hep-th/9704080.

\bibitem{hms}
G. Horowitz, J. Maldacena and A. Strominger, hep-th/9603109;
\plb {\bf 383} (1996) {151}.

\bibitem{gregory}
R. Gregory and R. LaFlamme, hep-th/9301052,
\prl {\bf 70} (1993) 2837.

\bibitem{halyo}
E. Halyo, hep-th/9705107.

\bibitem{cocv}
M. Costa and M. Cvetic, hep-th/9703204.

\bibitem{hp}
G. Horowitz and J. Polchinski, hep-th/9612146, 
\prd {\bf 55} (1997) {6189};
hep-th/9707170.

\bibitem{matstr}
{L. Motl, hep-th/9701025.}
{T. Banks and N. Seiberg, hep-th/9702187.}
{P.-M. Ho and Y.-S. Wu, hep-th/9703016.}
{R. Dijkgraaf, E. Verlinde, and H. Verlinde, hep-th/9703030.}
{Y. Imamura, hep-th/9703077.}

\bibitem{lmwip}
M. Li and E. Martinec, work in progress.


\end{thebibliography}
\end{document}